%% file: samplepaper.tex
\documentclass[article,xcolor=table]{llncs}
\usepackage{graphicx}
\usepackage{amsmath,amsthm,amssymb}

\usepackage[table]{xcolor}

\usepackage[sorting=none]{biblatex} 
\usepackage{xcolor}

\usepackage{wrapfig}

\usepackage{multirow}
\usepackage{rotating}
\usepackage{verbatim} 
\usepackage{booktabs}
\usepackage{graphicx}
\usepackage{caption}
\usepackage{subcaption}

\usepackage{parskip}

\usepackage{wrapfig}

\usepackage{algorithmic,algorithm}

\usepackage[breaklinks=true,hidelinks]{hyperref} 

\usepackage{svg}
\newcommand{\orcid}[1]{\href{https://orcid.org/#1}{\includegraphics[width=10pt]{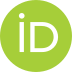}}}

\begin{document}   
\title{Early Experiences Migrating CUDA codes to oneAPI}
%
%
\author{Manuel Costanzo\inst{1}\orcid{0000-0002-6937-3943}
\and Enzo Rucci\inst{1}\orcid{0000-0001-6736-7358}\thanks{Corresponding author.} \and
Carlos García-Sánchez\inst{2}\orcid{0000-0002-3470-1097}  \and
Marcelo Naiouf\inst{1}\orcid{0000-0001-9127-3212}}
%
%
\institute{III-LIDI, Facultad de Informática, UNLP – CIC. \\La Plata (1900), Bs As, Argentina \\
\email{\{mcostanzo,erucci,mnaiouf\}@lidi.info.unlp.edu.ar} \\
 \and
Dpto. Arquitectura de Computadores y Automática, Universidad Complutense de Madrid.  Madrid (28040), España \\
\email{garsanca@dacya.ucm.es}}
\maketitle              
\begin{abstract}

The heterogeneous computing paradigm represents a real programming challenge due to the proliferation of devices with different hardware characteristics. Recently Intel introduced oneAPI, a new programming environment that allows code developed in DPC++ to be run on different devices such as CPUs, GPUs, FPGAs, among others. This paper presents our first experiences in porting two CUDA applications to DPC++ using the oneAPI \texttt{dpct} tool. From the experimental work, it was possible to verify that \texttt{dpct} does not achieve 100\% of the migration task; however, it  performs most of the work, reporting the programmer of possible pending adaptations. Additionally, it was possible to verify the functional portability of the DPC++ code obtained, having successfully executed it on different CPU and GPU architectures.

\end{abstract}
\input{intro}

\input{back-relwork}
\input{impl-res}
\input{conclusions}


%
%
%

\printbibliography
\end{document}

%% file: intro.tex
\section{Introduction}
\label{sec:intro}


In the last decade, the quest to improve the energy efficiency of computing systems has fueled the trend toward heterogeneous computing and massively parallel architectures~\cite{Giefers2016}. 
One effort to face some of the programming issues related to
heterogeneous computing is SYCL~\footnote{\url{https://www.khronos.org/registry/SYCL/specs/sycl-2020/pdf/sycl-2020.pdf}}, a new open standard from Khronos Group. SYCL is a domain-specific embedded language that allows the programmer to write single-source C++ host code
including accelerated code expressed as functors. In addition, SYCL features asynchronous task graphs, 
buffers defining location-independent storage, automatic overlapping kernels and communications, interoperability with OpenCL, among other characteristics~\cite{EarlyExperimentsUsingSYCL-FPGA}. 

Recently, Intel announced the \textit{oneAPI} programming ecosystem that provides a unified programming model for a wide range of hardware architectures.  At the core of the oneAPI environment is the Data Parallel C++ (DPC++) programming language, which can be summarized as C++ with SYCL. Additionally, DPC++ also features some vendor-provided extensions that might be integrated into these standards in the future~\cite{PortingLegacyCUDAtoOneAPI}. 


Today, GPUs can be considered the dominant accelerator and CUDA is the most popular programming language for them~\cite{CostanzoSpringerCACIC2020}. To tackle CUDA-based legacy codes, oneAPI provides a compatibility tool (\texttt{dpct}) that facilitates the migration to the SYCL-based DPC++ programming language. In this paper, we present our experiences from porting two original CUDA apps to DPC++ using \texttt{dpct}. Our contributions are: (1) the analysis of the \texttt{dpct} effectiveness for CUDA code migration, and (2) the analysis of the DPC++ code's portability, considering different target platforms (CPU and GPUs).

%% file: back-relwork.tex
\section{The oneAPI Programming Ecosystem}
\label{sec:back}

oneAPI~\footnote{\url{https://www.oneapi.com/}} is an industry proposal based on standard and open specifications, that includes the DPC++ language and a set of domain libraries. 
Each hardware vendor provides its  own compatible implementations targeting different hardware platforms, like CPUs and accelerators. The Intel oneAPI implementation consists of the Intel DPC++ compiler, the Intel \texttt{dpct} tool, multiple optimized libraries, and advanced analysis and debugging tools~\cite{ParallelUniverseOneAPI}.

%% file: impl-res.tex
\section{Experimental Work and Results}
\label{sec:imps}

\subsection{Migrating CUDA Codes to oneAPI}


\texttt{dpct} assists developers in porting CUDA code to DPC++, generating human readable code wherever possible. Typically,  \texttt{dpct} migrates ~80-90\% of code in automatic manner. In addition, inline comments are provided to help developers finish migrating the application. In this work, we have selected two CUDA applications  from the CUDA Demo Suite (CDS)~\footnote{\url{https://docs.nvidia.com/cuda/demo-suite/index.html}}. Both codes were translated from CUDA to DPC++ using the \texttt{dpct} tool.

\subsubsection{Matrix Multiplication (MM)}

\begin{figure}[t]
\centering
  \includegraphics[width=.8\linewidth]{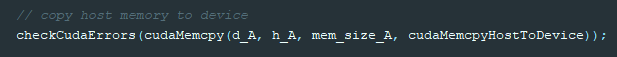}
\\
  \includegraphics[width=.8\linewidth]{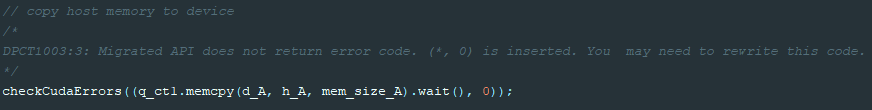}
\caption{MM memory transference. Up: Original CUDA code. Down: Resultant DPC++ code.}
\label{fig:mm-dpct}
\end{figure}

This app computes a MM using shared memory through tiled approach. Fig.~\ref{fig:mm-dpct} shows an example of the memory transference translations. Because \textit{checkCudaErrors} is a utility function (it is not part of the CUDA core), \texttt{dpct} inserts a comment to report this situation. Then, the programmer must decide whether to remove the function or redefine it.

\begin{figure}[t]
\centering
  \includegraphics[width=.8\linewidth]{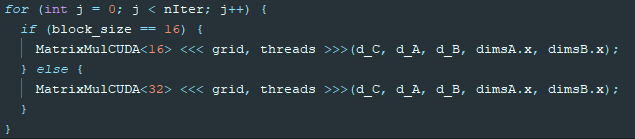}
\\
  \includegraphics[width=.8\linewidth]{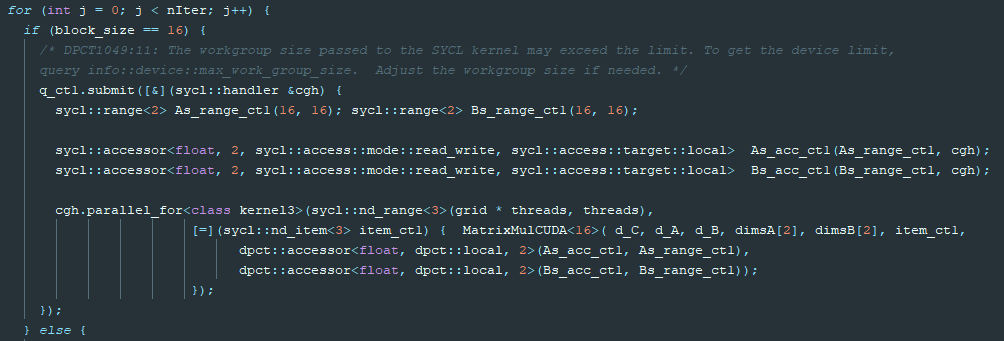}
\caption{MM kernel call. Up: Original CUDA code. Down: Resultant DPC++ code (portion).}
\label{fig:mm-kernel}
\end{figure}

Fig.~\ref{fig:mm-kernel} shows the kernel invocations. At the top, the original CUDA kernel's call and, al the bottom, the migrated DPC++ code (only a portion is included due to the lack of space). On the one hand, \texttt{dpct} adds comments informing the programmer that it is possible that the size of the \textit{work-group} exceeds the maximum of the device, being his responsibility to prevent this from happening. On the other hand, the resulting code is longer and more complex than the CUDA original code. However, it is important to remark that this code is the result of an automatic translation. By following the DPC++ conventions, it could be significantly simplified.

\begin{figure}[t]
\centering
\begin{subfigure}{.45\textwidth}
  \centering
  \includegraphics[width=.95\linewidth]{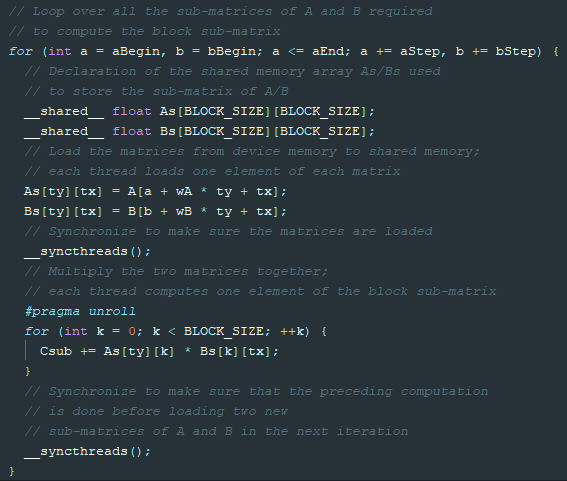}
\end{subfigure}%
\begin{subfigure}{.45\textwidth}
  \centering
  \includegraphics[width=.98\linewidth]{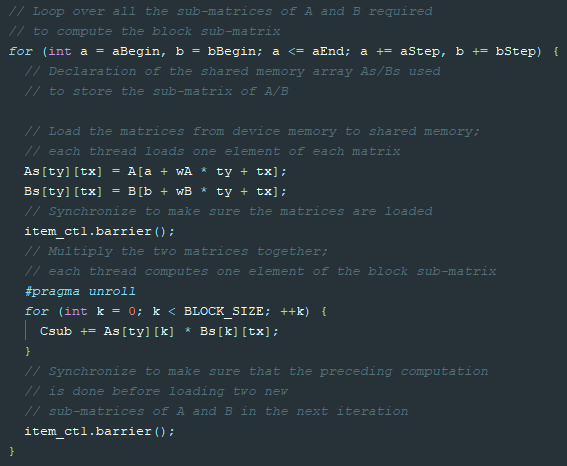}
\end{subfigure}
\caption{MM kernel. Left: Original CUDA code. Right: Resultant DPC++ code.}
\label{fig:mm-kernel-body}
\end{figure}

Finally, Fig.~\ref{fig:mm-kernel-body} shows part of the kernel bodies, resulting in very similar codes. \texttt{dpct} manages to correctly translate the local memory usage, although it defines the arrays outside the loop as opposed to the CUDA case. In addition, it can be noted that \texttt{dpct} effectively translates the \texttt{unroll} directive and the synchronization barriers.

\subsubsection{Reduction (RED)}

\begin{figure}[t]
\centering
  \includegraphics[width=.6\linewidth]{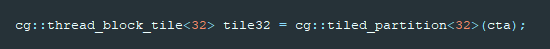}
\\
  \includegraphics[width=.8\linewidth]{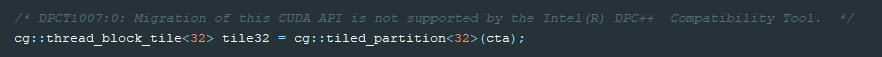}
\caption{RED kernel. Up: Original CUDA code. Down: Resultant DPC++ code.}
\label{fig:red}
\end{figure}

This app computes a parallel sum reduction of large arrays of values. The CUDA code includes several important optimization strategies like reduction using shared memory, \texttt{\_\_shfl\_down\_sync}, \texttt{\_\_reduce\_add\_sync} and \texttt{cooperative\_groups} reduce.

In this case, \texttt{dpct} is not able to translate advanced functionalities such as CUDA \textit{Cooperative Groups}. Fig.~\ref{fig:red} presents the comment inserted by \texttt{dpct} to inform the programmer about this issue. Even so, the tool manages to translate most of the original CUDA code, leaving little work to the programmer.

\subsection{Experimental Results}
\label{sec:results}

Two hardware platforms were used for the experimental work. The first comprises an Intel Core i3-4160 3.60GHz processor, 16GB main memory and a NVIDIA GeForce RTX 2070 GPU. The second has an Intel Core i9-10920X 3.50GHz processor, 32GB main memory, and an Intel Iris Xe MAX Graphics GPU, from the Intel DevCloud~\footnote{https://software.intel.com/content/www/us/en/develop/tools/devcloud.html}. oneAPI and CUDA versions are 2021.2 and 10.1, respectively. In addition, different workloads were configured for MM ($nIter=10$; $wA,wB,hA,hB=\{4096,8192,16384\}$). Finally, to run DPC++ code on NVIDIA GPUs, several modifications had to be made to the build, as it is not supported by default~\footnote{https://intel.github.io/llvm-docs/GetStartedGuide.html}. 

Table~\ref{tab:my-table} shows the execution times of MM (CUDA and DPC++ versions) on the different experimental platforms.
Before analyzing the execution times, it is important to remark that the DPC++ code was successfully executed on all the selected platforms and that the results were correct in all cases.

On the RTX 2070, the DPC++ code presents some overhead compared to the original code. However, it should be noted that these results are not final since the oneAPI support for NVIDIA GPUs is still experimental~\footnote{https://www.codeplay.com/portal/news/2020/02/03/codeplay-contribution-to-dpcpp-brings-sycl-support-for-nvidia-gpus.html}. In fact, currently the code generation does not consider any particular optimization passes.

The DPC++ code was compiled and successfully executed on two different Intel devices: a CPU and a GPU. In this way, we verified its functional portability on different architectures. Little can be said about its performance due to the absence of an optimized version for both Intel devices. However, there is probably significant room for improvement considering that the ported code was compiled and executed with minimal programmer intervention.

\begin{table}[t]
\centering
\caption{MM execution times on the target platforms}
\label{tab:my-table}
\resizebox{\textwidth}{!}{%
\begin{tabular}{p{0.1\textwidth} p{0.22\textwidth} p{0.22\textwidth} p{0.22\textwidth} p{0.22\textwidth}}

\toprule
\multicolumn{1}{c}{\textbf{Size}} & \multicolumn{1}{c}{\textbf{\begin{tabular}[c]{@{}c@{}}NVIDIA RTX 2070\\ (CUDA)\end{tabular}}} & \multicolumn{1}{c}{\textbf{\begin{tabular}[c]{@{}c@{}}NVIDIA RTX 2070\\ (oneAPI)\end{tabular}}} & \multicolumn{1}{c}{\textbf{Intel Core i9-10920X}} & \multicolumn{1}{c}{\textbf{Intel Iris Xe MAX}} \\ \midrule

4096 & 1.3 & 1.4 & 9.2 & 6.3 \\
8192 & 11.1 & 15.3 & 102.8 & 50.4  \\
16384 & 89.3 & 122.9 & 919.5 & 401.1 \\ \bottomrule
\end{tabular}%
}
\end{table}

%% file: conclusions.tex
\section{Conclusions and Future Work}
\label{sec:conc}

In this paper, we present our first experience migrating CUDA code to DPC++ using the Intel oneAPI enviroment. First, we were able to test the effectiveness of \texttt{dpct} for the selected test cases. Despite not translating 100\% of the code, the tool does most of the work, reporting the programmer of possible pending adaptations. Second, it was possible to verify the functional portability of the obtained DPC++ code, by successfully executing it on different CPU and GPU architectures.

As future work, we are interested in deepening the experimental work. In particular, we want to include other test cases, hardware architectures, and metrics (like performance portability).